\documentclass[journal]{IEEEtran}
\usepackage{amsmath}
\usepackage{graphicx}
\usepackage{algorithm} 
\usepackage{algpseudocode}
\usepackage[style=base]{caption,subcaption}

\usepackage{cuted}
\usepackage{multicol} 
\usepackage{booktabs}
\usepackage{amssymb}
\usepackage{color, soul}
\usepackage{cite}

\IEEEoverridecommandlockouts
\begin{document}
\title{Energy-Efficient Resource Allocation in Multi-UAV-Assisted Two-Stage Edge Computing for Beyond 5G Networks \thanks{}}

\author{Nway~Nway~Ei,~
	    Madyan~Alsenwi,~
	    Yan~Kyaw~Tun,~
	    Zhu Han,~\IEEEmembership{Fellow,~IEEE,}
	and~Choong~Seon~Hong,~\IEEEmembership{Senior~Member,~IEEE}
	
	\thanks{Nway Nway Ei, Madyan Alsenwi, Yan Kyaw Tun, and Choong Seon Hong  are with the Department of Computer Science and Engineering, Kyung Hee University,  Yongin-si, Gyeonggi-do 17104, Rep. of Korea, e-mail:{\{nwayei, malsenwi, ykyawtun7, cshong\}@khu.ac.kr}.}
	\thanks{Zhu Han is with the Electrical and Computer Engineering Department,
		University of Houston, Houston, TX 77004, and the Department of
		Computer Science and Engineering, Kyung Hee University, Yongin-si,
		Gyeonggi-do 17104,  Rep. of Korea, email{\{zhan2\}@uh.edu}.}}
	
\maketitle

\begin{abstract}
Unmanned aerial vehicle (UAV)-assisted multi-access edge computing (MEC) has become one promising solution for energy-constrained devices to meet the computation demand and the stringent delay requirement. In this work, we investigate a multiple UAVs-assisted two-stage MEC system in which the computation-intensive and delay-sensitive tasks of mobile devices (MDs) are cooperatively executed on both MEC-enabled UAVs and terrestrial base station (TBS) attached with the MEC server.
Specifically, UAVs provide the computing and relaying services to the mobile devices. In this regard, we formulate a joint task offloading, communication and computation resource allocation problem to minimize the energy consumption of MDs and UAVs by considering the limited communication resources for the uplink transmission, the computation resources of UAVs and the tolerable latency of the tasks. The formulated problem is a mixed-integer non-convex problem which is NP hard. Thus, we relax the channel assignment variable from the binary to continuous values. However, the problem is still non-convex due to the coupling among the variables. To solve the formulated optimization problem, we apply the Block Successive Upper-bound Minimization (BSUM) method which guarantees to obtain the stationary points of the non-convex objective function. In essence, the non-convex objective function is decomposed into multiple subproblems which are then solved in a block-by-block manner. Finally, the extensive evaluation results are conducted to show the superior performance of our proposed framework.      
	              	
\end{abstract}
\IEEEpeerreviewmaketitle

\section{Introduction}
\subsection{Background and Motivation}
With the unprecedented growth in the development of technology, the functionalities of smart devices such as smartphones, Internet of Things (IoT) devices, etc., have become more advanced. Moreover, the applications running on them for online gaming, augmented reality (AR), virtual reality (VR), video streaming and infotainment require high traffic demand and generate more processing data. This, in turn, leads to the requirement of more communication and computing related resources for such resource-constrained devices. Therefore, the cloud computing that provides the computing resources as well as the storage space has been introduced as a promising paradigm to lessen the burdens on mobile devices \cite{Mao2017}. However, offloading the tasks of mobile devices to the central cloud server that is generally distant from them incurs high latency and degrades the system efficiency.
To address this problem, multi-access edge computing (MEC) system that brings the computing resources near to the devices has been further introduced \cite{Mao2017}. In particular, by providing a distributed computing environment, the cloud servers that are installed at the edge of the network such as the access points or base stations mitigate the energy consumption and communication/computation delay experiencing at the mobile devices \cite{Hu2019}. However, constructing the new terrestrial network infrastructures in the temporary events (such as football matches or concerts) or in the disaster areas might not be cost-effective merely to assist the existing terrestrial network.

Recently, the use of unmanned aerial vehicles (UAVs) such as balloons, airships or drones as the communication and computing platforms has drawn much attention to the researchers. Due to the flexibility of on-demand deployment for the temporary events or emergency situations, UAVs are generally deployed as the assistance of the existing terrestrial networks in order to fulfill the unprecedented traffic demand and to provide the global internet connectivity. It is also anticipated that they are being used in various applications to bring fruitful business opportunities in the upcoming years \cite{zeng2019accessing}\cite{zeng2016wireless}.
Leveraging the good attributes of UAVs such as on-demand deployment and cost effectiveness, MEC-enabled UAVs can be deployed as the aerial computing platforms to offer the computing services to the energy-constrained mobile devices which are generally unable to completely execute the computation-intensive and delay-sensitive tasks locally.  With this approach, the devices can prolong their battery life as well as the overall system efficiency can be enhanced. In addition, since UAVs can establish the reliable line-of-sight communication link with the ground terminals, they can relay the tasks of the mobile devices to the TBS when the MDs cannot directly offload their tasks to the TBS due to the severe link blockage or poor channel condition.
\subsection{Challenges and Contributions}
When UAVs are considered as the edge computing platforms, it is challenging to determine the amount of tasks to be offloaded from mobile devices (MDs) to the UAVs and decide the optimal allocation of communication and computation resources of UAVs to their associated devices in an energy-efficient manner. Moreover, as for UAV being an energy-limited and resource-constrained device, it is hard to accomplish all the tasks offloaded from the MDs. To address that problem, we propose a multi-UAV-assisted two-stage MEC system in which MEC-enable UAVs and TBS cooperatively execute the offloaded tasks of the mobile devices. In particular, UAVs locally compute part of devices' offloaded tasks and relay the rest to the terrestrial base station (TBS) which has rich computing resources. Here, we assume that the devices cannot directly offload their tasks to the terrestrial base stations due to severe link blockage or poor channel condition. The main contributions of this paper are as follows:
\begin{itemize}
	\item Firstly, we investigate a multi-UAV-assisted two-stage MEC system in which multiple MEC-enabled UAVs offer the computing and relaying services to the mobile devices. Particularly, UAVs execute a partial portion of the tasks offloaded from the associated mobile devices according to their computing capacity and relay the rest of the tasks to TBS for further computing.  
	\item We then propose a joint resource allocation and task offloading problem in order to minimize the energy consumption of mobile devices and UAVs. The proposed problem is a mixed integer non-convex problem which is NP hard. 
	\item We further relax the channel allocation variable into the continuous form and then derive the upper bound of our objective function. To solve the formulated problem, we apply the block successive upper bound minimization (BSUM) algorithm which can tackle non-convex and non-smooth optimization problems.
	\item Finally, we perform an extensive simulation to verify that our proposed approach can yield the better solution compared to the other baseline schemes, namely, equal offloading, local processing only and offloading all.       
\end{itemize}  

The remainder of this paper is as follows. We describe the recent works in Section \ref{RecentWorks}. In Section \ref{sec: SystemModel}, we present our system model in detail. The communication and computation models for the proposed MEC system are described in Sections \ref{sec: CommunicationModel} and \ref{sec: ComputationModel}, respectively. Then, our formulated optimization problem and the proposed solution approach are provided in Section \ref{sec:ProblemFormulation}. The simulation results are illustrated in Section \ref{sec: SimulationResults}. Finally, we conclude the paper in Section \ref{sec: Conclusion}.      
 
\section{Recent Works} \label{RecentWorks}
\subsection{UAV-enabled Wireless Network}
The authors in \cite{Mozaffari2016a} analyzed the coverage and rate performance of a single UAV network underlying device-to-device communication links. They derived the average coverage probabilities for downlink users served by UAVs and for D2D users to show how the altitude of UAV and density of D2D users impact on the overall system. Exploiting the circle packing theory, the work in \cite{mozaffari2016efficient} addressed the problem of three-dimensional deployment of multiple aerial base stations in order to maximize the downlink coverage performance while serving ground users. In \cite{Sharma2016, mozaffari2017wireless, Mozaffari2016, Mozaffari2019}, the authors investigated how UAVs should be assigned to a certain area under the constraint of high traffic demand. Although UAVs can establish line-of-sight connections to the ground users promoting the network throughput, optimal on-demand deployment of multiple UAVs in an area where users are unevenly distributed is one of the challenging issues. Hence, the authors in \cite{Zhao2018} proposed two different deployment algorithms, centralized and distributed motion control algorithms to determine the minimum number of UAVs required for providing on-demand coverage with seamless connectivity.

Since UAVs are energy-constrained devices, the energy-efficient deployment and resource allocation is another challenging issue to be addressed. The authors in \cite{Mozaffari2017} introduced a multi-UAV assisted IoT network in which the transmit power of IoT devices is minimized by jointly optimizing the deployment and mobility of UAVs. In \cite{Ei2019}, the authors addressed the transmit power minimization problem in the UAV-assisted wireless network by jointly optimizing the altitude and transmit power of UAVs. The work in \cite{Zhang2019} investigated the problem of optimal subchannel allocation and UAV speed control to boost the uplink system sum-rate in a multi-UAV relays network. Predicting the content request distribution and mobility patterns of users with the help of the conceptor-based echo state network, the authors in \cite{Chen2017} conducted the proactive deployment of cache-enabled UAVs to maximize the quality of experience of users while reducing UAVs' transmit power consumption.   

\subsection{Multi-access Edge Computing}
Regarding to MEC, there are many existing works that addressed the two main problems of latency and energy utilization minimization. The authors in \cite{Ning2019} investigated the latency minimization problem in a cloud computing and MEC cooperated system for the partial offloading scenario. The work in \cite{Ren2017} studied the partial offloading scheme in a multi-user mobile edge computing system in which they minimized the weighted-sum latency of all users' devices by considering the communication and computation resource constraints. In \cite{Liu2017}, the authors minimized the long-term time average power consumption of user equipment by taking into account the constraints of delay and reliability. The optimal allocation of computing resources and task offloading policy is determined by applying the Lyapunov stochastic optimization method. While considering the task offloading in the MEC system, the assignment of tasks offloaded from the devices to the edge server must be properly investigated since it has an impact on the processing efficiency of the system. Therefore, the authors in \cite{gao2020computation} proposed a two-stage computation offloading framework in which the execution delay of the tasks is minimized by formulating the aggregative game among multiple user equipment.       
\subsection{UAV-assisted Multi-access Edge Computing} 
Employing MEC-enabled UAV brings fruitful advantages over a typical MEC scenario. The work in \cite{Chen2020} studied a three-dimensional UAV-aided MEC system for computation offloading of mobile users. Introducing a proactive deep reinforcement learning scheme, in this work, the expected long-term computation performance of the network is maximized by modeling the stochastic game among mobile users. With the aim of minimizing the total energy consumption of UAV and ground user equipment, the works in \cite{Hu2019}, \cite{tun2020energy,Alsenwi2020,Zhang2019a} studied the problem of managing the allocation of communication and computation resources to the mobile devices and optimizing the trajectory of UAV. Similarly, the authors in \cite{Li2020} maximized the energy efficiency of UAV by optimally determining the offloading strategy and transmission power of ground users and UAV trajectory. The work in \cite{Zhang2020} investigated the response time minimization problem in an aerial MEC system in which MEC-enable UAV is deployed to serve a swarm of UAVs with the communication and computation resources. 

Most of the existing works have mainly focused on resource allocation and UAV trajectory optimization problem in a single UAV-assisted MEC system for maximizing the energy efficiency or minimizing the delay. The multiple UAV-aided or ground-air integrated MEC system has been less explored. Different from the existing works, in this paper, we propose a multi-UAV-assisted two-stage MEC system in which UAVs offer computing and relaying services to MDs to help execute their tasks.  
        
\section{System Model} \label{sec: SystemModel}
\begin{figure}[t]
	\centering
	\captionsetup{justification = centering}
	\includegraphics[width=\linewidth]{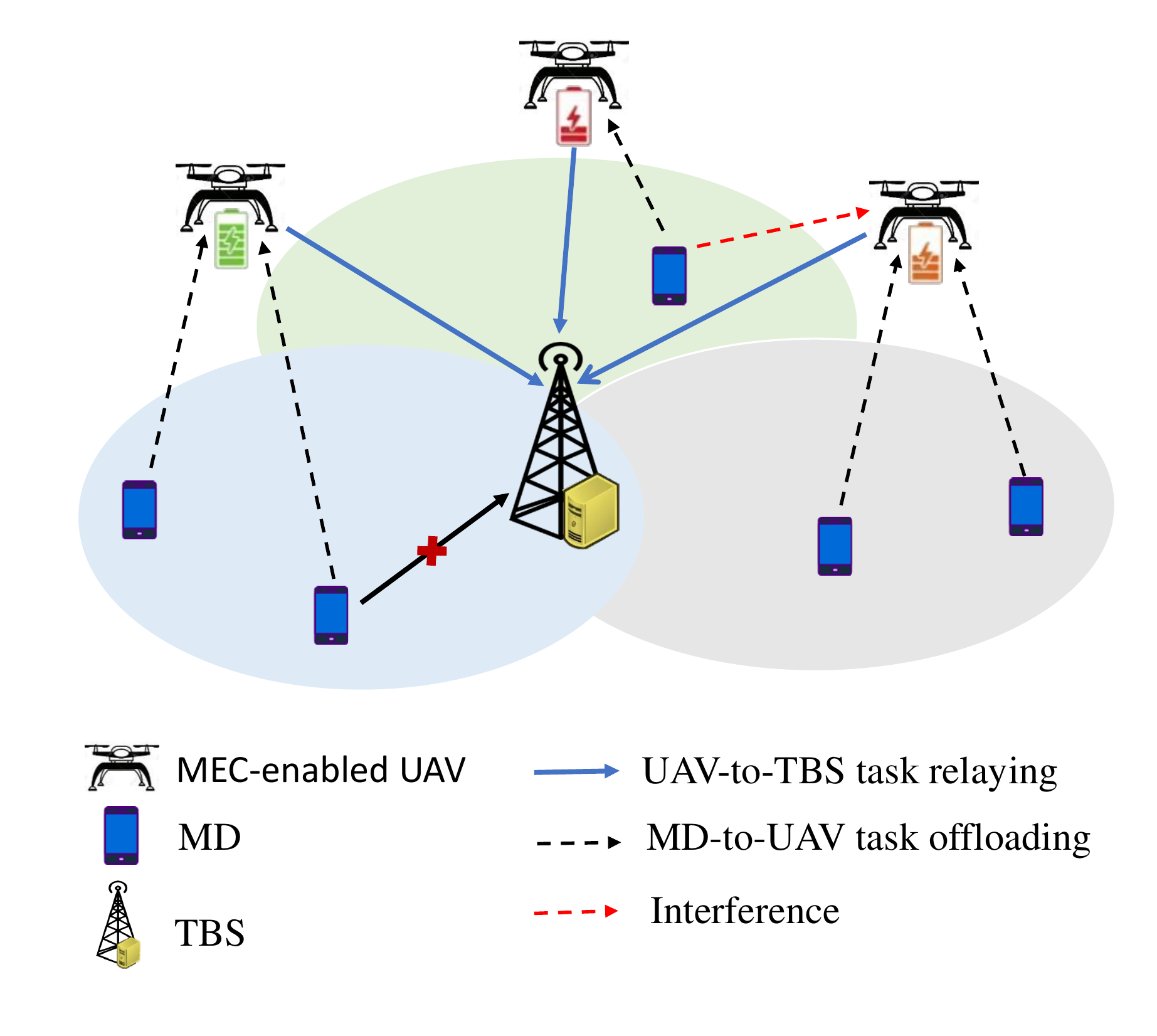}
	\caption{Multi-UAV-assisted MEC system.}
	\label{System Model}
\end{figure}
In our system model, as shown in Fig. 1, we propose a multi-UAV assisted two-stage MEC system in which there are $M$ MEC-enabled UAVs, a number of $U$ mobile devices and a terrestrial base station. A set $\mathcal{M}=\{1,2,...,M\}$ of MEC-enabled UAVs are deployed for providing computing and relaying services to a set of MDs $\mathcal{U}=\{1,2,...,U\}$ which are distributed in an area of interest. In this work, we assume that MDs cannot directly offload their tasks to the terrestrial base station due to the low signal strength or poor channel condition. Particularly, UAVs provide computing and relaying services to MDs for the execution of their tasks. Since UAVs are constrained by power and size, the available computing and communication resources on board are very limited. In that case, it is impossible for the UAVs to locally execute all the tasks offloaded from their associated MDs. The promising approach is that UAV can relay part of the MDs' offloaded tasks to the TBS which provides a high-speed transmission rate with grid power supply and is empowered with an ultra-high performance processing server. Leveraging the inherent attributes such as the ability to establish a line-of-sight communication links to the ground terminals and flexibly adjust the altitude, UAVs can provide not only the reliable communication but also the broader wireless coverage on the ground. Moreover, UAVs are assumed to be hovering or circling at the minimum fixed altitude enough to provide sufficient coverage without suffering severe path loss. 

Since the position of UAVs can significantly affect the network performance, we exploit the k-means clustering algorithm for the deployment of UAVs and assignment of MDs to them. Specifically, MDs are grouped into different clusters according to the number of available UAVs which are assumed to be centroids of the clusters. Here, we assume that there is a central controller that has prior knowledge about the locations of MDs and controls the UAVs' position. Given a set of MDs, $\mathcal{U}$, the k-means clustering algorithm intends to partition MDs into $M$ clusters, $\mathcal{U}_1,\mathcal{U}_2,...,\mathcal{U}_M$, which are mutually exclusive and collectively exhaustive sets, i.e., $\mathcal{U}_a \cap \mathcal{U}_b  = \emptyset,a \neq b$ and  $\mathcal{U}_1 \cup \mathcal{U}_2 \cup ...\cup \mathcal{U}_M = \mathcal{U}$. Moreover, the path loss and transmit power between UAVs and MDs can also be reduced by minimizing the squared deviation of MD's distance from its cluster's centroid. Denoting the two-dimensional coordinates of MD $u$ and UAV $m$ as $s_u = (x_u, y_u)$ and $s_m = (x_m, y_m)$, respectively, where $u \in \mathcal{U}$ and $m \in \mathcal{M}$, the association between MDs and UAVs can be obtained by solving the problem below:
\begin{equation}
\min_{\{\mathcal{U}_1,...,\mathcal{U}_M\}} \sum_{m=1}^{M}\sum_{u \in \mathcal{U}_m}\left\lVert s_m-s_u\right\rVert^2. \label{kmeans}
\end{equation}
After determining the association of MDs to UAVs in (\ref{kmeans}), we present the mathematical representation of our proposed system model and problem formulation in the following subsections.
 
\begin{table*}[ht!]
	\caption{Summary of Key Notations}
\begin{center}
\begin{tabular}{|p{1cm}|p{7cm}||p{1cm}|p{7cm}|}
	\hline
	\textbf{Notation} & \textbf{Definition} & \textbf{Notation} & \textbf{Definition} \\ \hline \hline
	\ $\mathcal{M}$ & Set of UAVs, $|\mathcal{M}| = M $ & $t_{u,m}^{\textrm{local}}$ & Delay for local computing	  \\ \hline  
	\ $\mathcal{U}$ & Set of mobile devices, $|\mathcal{U}| = U$ & $t_{u,m}^e$ & Delay for computing MD $u$'s offloaded task at associated UAV $m$ \\ \hline
	\ $\mathcal{U}_m$ & Set of associated mobile devices with UAV $m$ & $t_{u,m,0}^{\textrm{off}}$ & Delay for relaying portion of MD $u$'s task to TBS by UAV $m$   \\ \hline
	\ $\mathcal{N}$ & Set of subchannels available for MD to UAV data transmission & $t_{u,m}^{\textrm{off},n}$ & Delay of MD $u$ to transmit $l_{u,m}^{\textrm{off}}I_{u,m}$ bits of data to UAV $m$ over subchannel $n$  \\ \hline
	\ $\delta_{u,m}^n$ & Subchannel assignment variable & $f_{u,m}$ & Computing capacity of MD $u$ associated with UAV $m$  \\ \hline
	\ $h_{u,m}^n$ & Channel gain between MD $u$ and UAV $m$ over subchannel $n$ & $f_{u,m}^e$ & CPU frequency of UAV $m$ allocated to compute portion of MD $u$'s task  \\ \hline
	\ $h_{m,0}$ & Channel gain between UAV $m$ and TBS & $f_m^e$ & Maximum computing capacity of UAV $m$  \\ \hline
	\ $h_0$ & Channel gain at reference distance of 1 m & $E_{u,m}^{\textrm{local}}$ & Energy consumption of MD $u$ for local computing  \\ \hline
	\ $d_{u,m}$ & Distance between MD $u$ and UAV $m$ & $E_{u,m}^{\textrm{off},n}$ & Energy consumption of MD $u$ for task offloading to UAV $m$ \\ \hline
	\ $d_{m,0}$ & Distance between UAV $m$ and TBS & $E_m^e$ & Energy consumption of UAV $m$ for edge computing  \\ \hline
	\ $\alpha$ & Path loss exponent & $E_{m,0}^{\textrm{off}}$ & Energy consumption of UAV $m$ for relaying tasks to TBS  \\ \hline
	\ $P_{u,m}^n$ & Transmit power of MD $u$ to UAV $m$ over subchannel $n$ & $E_m^{\textrm{hov}}$ & Hovering energy of UAV $m$  \\ \hline
	\ $P_{m,0}$ & Transmit power of UAV $m$ to TBS & $w$ & Bandwidth of subchannel $n$  \\ \hline
	\ $\gamma_{u,m}^n$ & SINR for MD $u$ associated with UAV $m$ on subchannel $n$ & $\beta_{m,0}$ & Bandwidth allocated to UAV $m$ for UAV to TBS data transmission  \\ \hline
	\ $\gamma_{m,0}$ & SNR for UAV $m$ - TBS transmission & $\zeta$ & Thrust that depends on mass of UAV  \\ \hline
	\ $N_0$ & Noise power spectral & $\eta_m$ & Power efficiency of UAV $m$  \\ \hline
	\ $R_{u,m}^n$ & Data rate of MD $u$ associated with UAV $m$ over subchannel $n$ & $q$ & Number of rotors in each UAV 	 \\ \hline
	\ $R_{m,0}$ & Data rate of the link between UAV $m$ and TBS & $r$ & Diameter of rotor  \\ \hline
	\ $I_{u,m}$ & Total task input data size of MD $u$ associated with UAV $m$ & $\rho$ & Air density  \\ \hline
	\ $O_{u,m}$ & Required computing resource to execute 1-bit of data & $\psi$ & Rounding threshold  \\ \hline
	\ $T_{u,m}$ & The maximum tolerable delay for the completion of task & $\Delta$ & Maximum violation value\\ \hline
	\ $l_{u,m}^{\textrm{off}}$ & Portion of task data size offloaded to UAV $m$ by MD $u$ & $\tau$ & Weight parameter  \\ \hline	
	\ $\phi_{u,m,0}$ & Portion of task data size relayed to TBS by UAV $m$ & $\vartheta$ & Penalty parameter \\ \hline
	\ $B$ & Total available bandwidth for UAV to TBS communication & $k, k'$ & Constants that depend on processor's chip architecture \\ \hline
\end{tabular}
\end{center}
\end{table*}
\subsection{Communication Model} \label{sec: CommunicationModel}
In order to manage the communication resources, we consider that the total available system bandwidth is orthogonally divided into two portions for the MD-to-UAV data transmission which is used for offloading tasks from MDs to UAVs and UAV-to-TBS data transmission which is reserved for relaying tasks from UAVs to TBS, respectively. Then, the total available bandwidth for MD-to-UAV data transmission is further  divided into $N$ subchannels, denoted by a set $\mathcal{N}=\{1,2,...,N\}$, each with a bandwidth of $w=180$ kHz. The subchannels are shared by the mobile devices while transmitting their tasks to the associated UAVs.  
\subsubsection{MD-to-UAV Data Transmission}
Each MD offloads its computation-intensive and delay-sensitive tasks to the associated UAV in order to consume less energy on local computation. To model the data transmission link between MDs and UAVs, we consider that the line-of-sight link is available and adopt the Rician channel fading model. In this work, we consider that the orthogonal frequency division multiple access (OFDMA) system is leveraged among MDs associated to each UAV to avoid intra-cell interference. We now define $\delta_{u,m}^n \in \{0,1\}$ as a subchannel assignment variable, which indicates whether or not subchannel $n$ is allocated to MD $u$ associated with UAV $m$ as follows:
\begin{equation}
\delta_{u,m}^{n}=
\begin{cases}
1, \ \ \text{if subchannel $n$ is assigned to MD $u$ to} \\ 
\ \ \ \  \text{transmit task data to UAV $m$},\\
0, \ \  \text{otherwise}.
\end{cases}
\end{equation}
Then, adopting the free-space path loss model, the channel gain between MD $u$ and UAV $m$ over subchannel $n$ is given in \cite{wu2018common}\cite{Al-Hourani2014}:

\begin{equation}
h_{u,m}^{n} = \frac{h_0}{(d_{u,m})^\alpha}, 
\end{equation} 
where $h_0$ is the channel gain at a reference distance of $1$m and $\alpha$ is the path loss exponent. $d_{u,m}$ is the euclidean distance between MD $u$ and UAV $m$, i.e., $d_{u,m} =  \sqrt{(x_m-x_u)^2+(y_m-y_u)^2+(z_m-z_u)^2}$, where $(z_m-z_u)$ means the vertical distance between MD $u$ and UAV $m$.
Then, the signal to interference plus noise ratio (SINR) for MD $u$ associated with UAV $m$ over subchannel $n$ is expressed as follow:
\begin{equation}
\gamma_{u,m}^n = \frac{P_{u,m}^n h_{u,m}^n}{\sum_{\substack{m' \in \mathcal{M},\\ m' \ne m}}\sum_{\substack{u' \in \mathcal{U}, \\ u' \ne u}}\delta_{u',m'}^n P_{u',m'}^nh_{u',m'}^n + N_0},
\end{equation}
where $P_{u,m}^n$ is the transmit power of MD $u$ on subchannel $n$ and $N_0$ is the noise power spectral. Here, we take into account the interference from other mobile devices associated with UAV $m' \in \mathcal{M}, m'\ne m$ transmitting on the same subchannel $n$.
Hence, the achievable data rate for MD $u$ that is associated with UAV $m$ on subchannel $n$ is given by
\begin{equation}
R_{u,m}^n = w\log_2\left(1 + \gamma_{u,m}^n\right), 
\end{equation}
where $w$ is the bandwidth of subchannel $n$.

\subsubsection{UAV-to-TBS Data Transmission}
We assume that MDs cannot directly offload their tasks to the TBS due to the severe link blockage or poor channel condition. Moreover, UAVs are constrained by their computing capacity to fulfill the computation demand from the MDs. Hence, UAVs will further relay a partial portion of the MDs' offloaded tasks to the TBS in addition to local execution. The communication link between UAVs and TBS is also assumed to be dominated by the line-of-sight link as in MD-to-UAV data transmission. The channel gain between UAV $m$ and TBS which is located at $(x_0,y_0,z_0)$ is given by 
\begin{equation}
h_{m,0} = \frac{h_0}{(d_{m,0})^\alpha},  
\end{equation}
where $d_{m,0}=\sqrt{(x_m-x_0)^2+(y_m-y_0)^2+(z_m-z_0)^2}$ is the distance between UAV $m$ and TBS. 

For the transmission link between UAVs and TBS, we consider that the available bandwidth $B$ is proportionally allocated to $M$ UAVs so that there is no interference among them when they relay MDs' offloaded tasks to the TBS. Therefore, the signal to noise ratio (SNR) of UAV $m$ for relaying task data to the TBS can be calculated as
\begin{equation}
\gamma_{m,0} = \frac{P_{m,0} h_{m,0}}{N_0},
\end{equation}
where $P_{m,0}$ the transmit power of UAV $m$ to the TBS.
The data rate achieved by UAV $m$ for transmission to the TBS is given by 
\begin{equation}
R_{m,0} = \beta_{m,0} \log_2\left(1+\gamma_{m,0}\right),
\end{equation}
where $\beta_{m,0}=\frac{B}{M}$ is the bandwidth allocated to UAV $m$ for communication with the TBS. 

\subsection{Computation Model} \label{sec: ComputationModel}
Let us suppose the computing task of MD $u$ associated with UAV $m$ is denoted as a tuple $\left(I_{u,m}, O_{u,m}, T_{u,m}\right)$, where $I_{u,m}$ is the data size of the computing task, $O_{u,m}$ is the amount of required computing resources to execute 1-bit of input data and $T_{u,m}$ denotes the maximum tolerable delay for the completion of task. Each mobile device is assumed to be able to perform local computing and computation offloading simultaneously.

\subsubsection{Local Computing at MD} 
Since each MD has very limited energy and computing resources, it is impossible to complete the tasks in time if it only relies on the local computing. Hence, we consider that MD $u$ offloads $l_{u,m}^{\textrm{off}}I_{u,m}$ (in bits) to UAV $m$, where $l_{u,m}^{\textrm{off}} \in [0,1]$ and computes the amount of task $\left(1 - l_{u,m}^{\textrm{off}}\right)I_{u,m}$ locally. It is assumed that the task transmission and computing at the mobile device can be done in a simultaneous manner \cite{Hu2019}. The time taken for MD $u$ associated with UAV $m$ to compute the task locally is given by
\begin{equation}
t_{u,m}^{\textrm{local}} = \frac{\left(1 - l_{u,m}^{\textrm{off}}\right)I_{u,m}O_{u,m}}{f_{u,m}},
\end{equation}
where $f_{u,m}$ is the CPU frequency of MD $u$ associated with UAV $m$.

The energy consumption of MD $u$ associated with UAV $m$ can be expressed as \cite{Hu2019},
\begin{equation}
E_{u,m}^{\textrm{local}} = \left(1 - l_{u,m}^{\textrm{off}}\right)I_{u,m}O_{u,m}kf_{u,m}^2,
\end{equation}
where $k$ is the constant that depends on the processor's chip architecture.

\subsubsection{Computation offloading to UAV and TBS}
When MD $u$ offloads tasks to its associated UAV $m$ over subchannel $n$ for remote computing, the delay of MD $u$ is calculated by
\begin{equation}
t_{u,m}^{\textrm{off},n} = \frac{l_{u,m}^{\textrm{off}}I_{u,m}}{w\log_2\left(1 + \gamma_{u,m}^n\right)}.
\end{equation}

The energy consumption of MD $u$ when it transmits the task to its associated UAV $m$ over subchannel $n$ is given by 
\begin{equation}
E_{u,m}^{\textrm{off},n} = P_{u,m}^nt_{u,m}^{\textrm{off},n} =   \frac{l_{u,m}^{\textrm{off}}I_{u,m}P_{u,m}^n}{w\log_2\left(1 + \gamma_{u,m}^n\right)}.
\end{equation}

UAV itself is an energy and resource constrained device, it cannot handle all the tasks offloaded from its associated MDs. Therefore, UAVs collaborate with the TBS to reduce the computation burden on them. In essence, each UAV executes a portion of MD $u$'s offloaded tasks locally and relays the rest to the TBS to save its energy consumption as well as computing resources. Hence, the time taken for UAV $m$ to compute portion of MD $u$'s offloaded task is calculated as
\begin{equation}
t_{u,m}^e=\frac{\left(1-\phi_{u,m,0}\right)l_{u,m}^{\textrm{off}}I_{u,m}O_{u,m}}{f_{u,m}^e},
\end{equation}
where $f_{u,m}^e$ is the CPU frequency of UAV $m$ allocated to MD $u$ for the task execution and $\phi_{u,m,0}\in [0,1]$ denotes the portion of MD $u$'s offloaded task at UAV $m$ that will be further relayed to the TBS.

The energy consumed by UAV $m$ to partially compute the offloaded tasks of its associated MD $u$ is expressed as
\begin{equation}
E_{u,m}^e = \left(1-\phi_{u,m,0}\right)l_{u,m}^{\textrm{off}}I_{u,m}O_{u,m}k'(f_{u,m}^e)^2.
\end{equation}

Moreover, we consider that all the associated MDs of UAV offload their tasks, but the UAV does not have enough computing resources. Therefore, part of MD's offloaded task-input data at the UAV is assumed to be further relayed to the TBS. The latency incurred by UAV $m$ to relay MD $u$'s offloaded tasks to the TBS is calculated by
\begin{equation}
t_{u,m,0}^{\textrm{off}} = \frac{\phi_{u,m,0}l_{u,m}^{\textrm{off}}I_{u,m}}{\beta_{m,0} \log_2\left(1+\gamma_{m,0}\right)}.
\end{equation}

Then, the energy consumption of UAV $m$ when it transmits part of its associated MD $u$'s task-input data to the TBS can be calculated as 
\begin{equation}
E_{u,m,0}^{\textrm{off}} = P_{m,0}t_{u,m,0}^{\textrm{off}} =  \frac{\phi_{u,m,0}l_{u,m}^{\textrm{off}}I_{u,m}P_{m,0}}{\beta_{m,0}\log_2\left(1+\gamma_{m,0}\right)}.
\end{equation}

The total energy consumption of MD $u$ associated with UAV $m$ for local computing and task offloading can be expressed as 
\begin{equation}
E_{u,m}^{\textrm{tot}} = E_{u,m}^{\textrm{local}} + \sum_{n=1}^{N}\delta_{u,m}^nE_{u,m}^{\textrm{off},n}.
\end{equation}
Both MDs and UAVs are assumed to be able to perform computing and offloading the tasks simultaneously. It should be noted that UAV needs to hover at a fixed altitude over the area of interest until all of its associated MDs' tasks have finished completely. Therefore, the time taken by UAV $m$ to hover over the area while providing communication and computing services to its associated MDs is denoted as
\begin{equation}
t_m^{\textrm{hov}} = \max_{u \in \mathcal{U}_m}\left\{\sum_{n=1}^{N}\delta_{u,m}^nt_{u,m}^{\textrm{off},n} + \max\left(t_{u,m}^e, \ t_{u,m,0}^{\textrm{off}}\right)\right\}. 
\end{equation}
The power consumed by UAV $m$ to hover over the area of interest is given by \cite{Monwar2018}\cite{Stolaroff2018}: 
\begin{equation}
P_m^{\textrm{hov}} = \frac{\zeta\sqrt{\zeta}}{\eta_m \sqrt{0.5\pi qr^2\rho}}, 
\end{equation} 
where $\zeta$ is the thrust that depends on the mass of UAV, $\eta_m$ denotes the power efficiency of UAV $m$, and $q$ is the number of rotors in each UAV. $r$ and $\rho$ are the diameter of rotor and air density, respectively. Hence, the hovering energy consumption of UAV $m$ is calculated as
\begin{equation}
E_m^{\textrm{hov}} = P_m^{\textrm{hov}}t_m^{\textrm{hov}}.
\end{equation}

The total energy consumption of UAV $m$ for remote computing, offloading and hovering is expressed as 
\begin{equation}
E_m^{\textrm{tot}} = E_m^{\textrm{hov}} + \sum_{u=1}^{|\mathcal{U}_m|}\left(E_{u,m}^e + E_{u,m,0}^{\textrm{off}}\right).
\end{equation}
The total time taken for device $u$ to finish the task can be expressed as
\begin{equation}
t_{u,m} = \max\left[t_{u,m}^{\mathsf{local}}, \ \sum_{n=1}^{N}\delta_{u,m}^nt_{u,m}^{\mathsf{off},n} + \max\left(t_{u,m}^e,\ t_{u,m,0}^{\mathsf{off}}\right)\right].
\end{equation}
\section{Problem Formulation and Solution Approach} \label{sec:ProblemFormulation}
\subsection{Problem Formulation}
 In this section, we present our proposed joint task offloading, communication and computation resource allocation problem. The objective is to minimize the energy consumption of mobile devices and UAVs in the system and the optimization problem is formulated as follows:
\begin{align*}
 \underset{\boldsymbol{\delta,l,f,\phi}}{\min} \ 
 & \sum_{m=1}^{M}\sum_{u=1}^{|\mathcal{U}_m|}\left(E_{u,m}^\mathsf{local} + \sum_{n=1}^{N}\delta_{u,m}^nE_{u,m}^{\mathsf{off},n}\right) + \sum_{m=1}^{M}E_m^{\mathsf{tot}} \tag{23} \label{eq23}\\
 \text{s.t.} \\
 & t_{u,m} \le T_{u,m}, \forall u \in \mathcal{U}_m, \forall m \in \mathcal{M}, \tag{23a} \label{23a}\\
 & 0 < l_{u,m}^{\mathsf{off}} \le 1, \forall u \in \mathcal{U}_m, \forall{m} \in \mathcal{M}, \tag{23b}\\
 & \sum_{u=1}^{|\mathcal{U}_m|}f_{u,m}^e \le f_m^e,  \forall m \in \mathcal{M}, \tag{23c} \\
 & f_{u,m}^e \ge 0, \forall u \in \mathcal{U}_m, \forall{m} \in \mathcal{M}, \tag{23d}\\
 & 0 \le \phi_{u,m,0} \le 1, \forall u \in \mathcal{U}_m, \forall m \in \mathcal{M}, \tag{23e} \\
 & \sum_{n=1}^{N}\delta_{u,m}^n \le 1, \forall u \in \mathcal{U}_m, \forall m \in \mathcal{M}, \tag{23f} \label{23f}\\ 
 & \delta_{u,m}^n \in \{0,1\}, \forall u \in \mathcal{U}_m, \forall n \in \mathcal{N}, \forall
 m \in \mathcal{M}, \tag{23g} \label{eq23g}
\end{align*}
where $\boldsymbol{\delta}=\{\delta_{u,m}^n\}_{n \in \mathcal{N}, u \in \mathcal{U}_m, m \in \mathcal{M}}$, $\boldsymbol{l}=\{l_{u,m}^{\textrm{off}}\}_{u \in \mathcal{U}_m, m \in \mathcal{M}}$, $\boldsymbol{f}=\{f_{u,m}^e\}_{u \in \mathcal{U}_m, m \in \mathcal{M}}$ and $\boldsymbol{\phi}=\{\phi_{u,m,0}\}_{u \in \mathcal{U}_m, m \in \mathcal{M}}$.
Constraint (23a) ensures that the task of MD $u$ has completed during the tolerable amount of time. Constraint (23b) means that the offloaded task input data size is less than the total input data size. Constraints (23c) and (23d) guarantee that the total allocated computing resources to its associated mobile devices does not exceed the maximum computing capacity of each UAV. Constraints (23e) states that the task data size offloaded to the TBS is less than MD $u$'s offloaded task at UAV $m$. Constraints (23f) and (23g) ensure that an associated mobile device of each UAV can only be allocated at most one subchannel.

The formulated optimization problem in (\ref{eq23}) is a mixed-integer non-convex problem which cannot be solved in a polynomial-time due to its combinatorial complexity \cite{Hong2016a}. Moreover, the subchannel assignment variable and the existence of coupling among the variables make it more challenging to solve. Therefore, we apply the BSUM framework to solve our problem which is presented in Section \ref{sec: SolutionApproach}.

\subsection{BSUM-based Joint Resource Allocation and Offloading} \label{sec: SolutionApproach}
In this section, we present our solution approach to the non-convex problem in (\ref{eq23}). The following steps are summarized to achieve the optimal solution of the proposed problem:
	\begin{itemize}
		\item Firstly, we reformulate problem (\ref{eq23}) into (\ref{eq24}) by relaxing the channel assignment variable. 
		\item Then, we propose the upper bound approximation function of the relaxed problem (\ref{eq24}) in (\ref{eq26}).
		\item After that, instead of minimizing problem (\ref{eq24}), we minimize the approximation function in (\ref{eq26}).
		\item Finally, we apply the rounding technique to enforce the subchannel assignment variable to be binary value.     
	\end{itemize}

 First, our proposed problem is reformulated by relaxing the channel assignment variable $\delta_{u,m}^n$ in constraint (\ref{eq23g}) into a continuous form as follows:
\begin{align*}
 \underset{\boldsymbol{\delta}, \boldsymbol{l}, \boldsymbol{f}, \boldsymbol{\phi}}{\min} \ 
& \sum_{m=1}^{M}\sum_{u=1}^{|\mathcal{U}_m|}\left(E_{u,m}^\textrm{local} + \sum_{n=1}^{N}\delta_{u,m}^nE_{u,m}^{\textrm{off},n}\right) + \sum_{m=1}^{M}E_m^{\textrm{tot}} \tag{24} \label{eq24} \\
\text{s.t.} \\
& (\text{23a})-(\text{23f}), \tag{24a} \\
&\delta_{u,m}^n \in [0,1], \forall u \in \mathcal{U}_m, \forall n \in \mathcal{N}, \forall
m \in \mathcal{M}. \tag{24b}
\end{align*}

Then, to put our proposed problem into the framework of BSUM, the objective function in (\ref{eq24}) is rewritten in a simple form as 
\begin{equation}
\underset{\boldsymbol{\delta} \in \mathcal{D}, \boldsymbol{l} \in \mathcal{L}, \boldsymbol{f} \in \mathcal{F}, \boldsymbol{\phi} \in \Phi}{\min}
 \mathcal{E}(\boldsymbol{\delta}, \boldsymbol{l}, \boldsymbol{f}, \boldsymbol{\phi})   \ \ \ \tag{25} \label{eq25}
\end{equation}
where \\ $\mathcal{E}(\boldsymbol{\delta}, \boldsymbol{l}, \boldsymbol{f}, \boldsymbol{\phi}) \triangleq \sum_{m=1}^{M}\sum_{u=1}^{|\mathcal{U}_m|}\left(E_{u,m}^\textrm{local} + \sum_{n=1}^{N}\delta_{u,m}^nE_{u,m}^{\textrm{off},n}\right)\\ + \sum_{m=1}^{M}E_m^{\textrm{tot}}$ is the objective function with the feasible sets of $\boldsymbol{\delta}, \boldsymbol{l}, \boldsymbol{f},$ and $\boldsymbol{\phi}$ given below,
\begin{align*}
 \mathcal{D} \triangleq & \{{\boldsymbol{\delta}}: t_u \le T_{u,m}, \forall u \in \mathcal{U}_m, 
 \forall m \in \mathcal{M}, \sum_{n=1}^{N}\delta_{u,m}^n \le 1, \forall u \in \\  &\mathcal{U}_m, 
 \forall m \in \mathcal{M},\delta_{u,m}^n \in [0, 1], \forall u \in \mathcal{U}_m, \forall n \in \mathcal{N}, \forall m \in \\ &\mathcal{M}\},
\end{align*} 
\begin{align*}
\mathcal{L} \triangleq & \{\boldsymbol{l}: t_u \le T_{u,m}, \forall u \in \mathcal{U}_m, \forall m \in \mathcal{M}, 0 < l_{u,m}^{\textrm{off}} \le 1, \forall u \in \\ &\mathcal{U}_m, 
\forall m \in \mathcal{M}\},
\end{align*} 
\begin{align*}
\mathcal{F} \triangleq &\{\boldsymbol{f}: t_u \le T_{u,m}, \forall u \in \mathcal{U}_m, \forall m \in \mathcal{M},\sum_{u=1}^{|\mathcal{U}_m|}f_{u,m}^e \le f_m^e,  \forall m \\ & \in 
\mathcal{M},f_{u,m}^e \ge 0, \forall u \in \mathcal{U}_m, \forall{m} \in \mathcal{M}\},
\end{align*} 
\begin{align*} 
\Phi \triangleq &\{\boldsymbol{\phi}: t_u \le T_{u,m}, \forall u \in \mathcal{U}_m, \forall m \in \mathcal{M}, 0 \le \phi_{u,m,0} \le 1, \forall u \\ & \in  
\mathcal{U}_m, \forall m \in  \mathcal{M}\}.
\end{align*} 

The problem in (\ref{eq25}) is still non-convex due to the existence of coupling among the variables such as $\boldsymbol{l}$, $\boldsymbol{f}$ and $\boldsymbol{\phi}$. Hence, to address this problem, we exploit the BSUM algorithm, a general type of block coordinate descent (BCD) algorithm \cite{Hong2016}\cite{Ndikumana2020a}. Practically, BCD cannot be directly applied to solve non-convex problems and is hard to guarantee the convergence to the set of stationary points of the objective function. One of the key advantages of the BSUM algorithm over BCD is that it can provide a good approximate solution of a non-convex objective function enough for the algorithm keep going under the practical and theoretical considerations. Literally, it successively minimizes the upper-bound approximation function by updating the blocks of variables in turn and can guarantee a few descent of the original objective function.     

Here, we define the convex surrogate function $\tilde{\mathcal{E}}_i(\boldsymbol{\delta}, \boldsymbol{l}, \boldsymbol{f},\boldsymbol{\phi})$ by adding the quadratic penalty term to the objective function and it can be described as
\begin{align}
\begin{split}
\tilde{\mathcal{E}}_i(\boldsymbol{\delta}_i; \boldsymbol{\delta}^{(r)}, \boldsymbol{l}^{(r)}, \boldsymbol{f}^{(r)}, \boldsymbol{\phi}^{(r)})&:=   
\mathcal{E}(\boldsymbol{\delta}_i; \boldsymbol{\tilde{\delta}}, \boldsymbol{\tilde{l}}, \boldsymbol{\tilde{f}}, \boldsymbol{\tilde{\phi}})+\frac{\vartheta_i}{2}\\     
& \left\lVert(\boldsymbol{\delta}_i - \boldsymbol{\tilde{\delta}})\right\rVert^2,   
\end{split} \ \ \ \tag{26} \label{eq26}
\end{align}
where $\vartheta_i$ is the positive penalty parameter.

Given the initial feasible points $\boldsymbol{\tilde{\delta}}$, $\boldsymbol{\tilde{l}}$, $\boldsymbol{\tilde{f}}$, and $\boldsymbol{\tilde{\phi}}$, instead of minimizing the intractable problem in (\ref{eq25}), we minimize the surrogate function in (\ref{eq26}) by separating into blocks. It is noted that the problem in (\ref{eq26}) is strictly convex because of the quadratic penalty term \cite{Boyd2004}. Let us suppose $i \in \mathcal{B}^r$, where $\mathcal{B}^r$ is the set of index blocks at iteration $r$. The similar approach can be applied for other variable blocks $\boldsymbol{l}$, $\boldsymbol{f}$, and $\boldsymbol{\phi}$. At each iteration $r+1$, we solve the following optimization problems to get the optimal solution of (\ref{eq26}),
\begin{align*}
&\boldsymbol{\delta}_i^{(r+1)} \in \min_{\boldsymbol{\delta}_i \in \mathcal{D}}\tilde{\mathcal{E}}_i\left(\boldsymbol{\delta}_i; \boldsymbol{\delta}^{(r)}, \boldsymbol{l}^{(r)}, \boldsymbol{f}^{(r)}, \boldsymbol{\phi}^{(r)}\right), \tag{27}  \label{eq27}\\
&\boldsymbol{l}_i^{(r+1)} \in \min_{\boldsymbol{l}_i \in \mathcal{L}}\tilde{\mathcal{E}}_i\left(\boldsymbol{l}_i; \boldsymbol{\delta}_i^{(r+1)}, \boldsymbol{l}^{(r)}, \boldsymbol{f}^{(r)}, \boldsymbol{\phi}^{(r)}\right), \tag{28} \label{eq28}\\
&\boldsymbol{f}_i^{(r+1)} \in \min_{\boldsymbol{f}_i \in \mathcal{F}}\tilde{\mathcal{E}}_i\left(\boldsymbol{f}_i; \boldsymbol{\delta}_i^{(r+1)}, \boldsymbol{l}_i^{(r+1)}, \boldsymbol{f}^{(r)}, \boldsymbol{\phi}^{(r)}\right), \tag{29} \label{eq29}\\
&\boldsymbol{\phi}_i^{(r+1)} \in \min_{\boldsymbol{\phi}_i \in \Phi}\tilde{\mathcal{E}}_i\left(\boldsymbol{\phi}_i; \boldsymbol{\delta}_i^{(r+1)}, \boldsymbol{l}_i^{(r+1)}, \boldsymbol{f}_i^{(r+1)}, \boldsymbol{\phi}^{(r)}\right). \tag{30} \label{eq30}
\end{align*}

Since the solution of the relaxed problem in (\ref{eq24}) cannot guarantee the subchannel assignment variable, $\delta_{u,m}^n$ to be binary value, the rounding technique is adopted to enforce the binary value of $\delta_{u,m}^n$ \cite{feige_et_al:LIPIcs:2016:6631}\cite{ndikumana2020deep}. Let suppose the rounding threshold be $\psi \in (0, 1)$. The optimal subchannel assignment value, $\delta_{u,m}^{n^*}$  is determined as follow:
\begin{equation}
\delta_{u,m}^{n^*} =
\begin{cases}
1, \ \ \text{if $\delta_{u,m}^{n^*} \geq \psi$,} \\ 
0, \ \ \text{otherwise}. \tag{31}
\end{cases}
\end{equation}
To address the problem of violating the communication resource constraint, we solve $\tilde{\mathcal{E}}_i + \tau\Delta$ by modifying the communication constraint in (\ref{23f}) by 
\begin{equation}
 \sum_{n=1}^{N}\delta_{u,m}^n \le 1 + \Delta, \forall u \in \mathcal{U}_m, \forall m \in \mathcal{M}, \tag{32}
\end{equation} 
where $\Delta$ is the maximum violation of the communication constraint and $\tau$ is the weight parameter of $\Delta$. Then, the value of $\Delta$ is expressed as
\begin{equation}
\Delta = \max\left\{0, \sum_{n=1}^{N}\delta_{u,m}^n - 1\right\}, \forall u \in \mathcal{U}_m, \forall m \in \mathcal{M}.  \tag{33}
\end{equation}
Using the value of $\Delta$ and solving $\tilde{\mathcal{E}}_i(\boldsymbol{\delta_i}^*,\boldsymbol{l_i}^*, \boldsymbol{f_i}^*, \boldsymbol{\phi_i}^*) + \tau\Delta$, we can obtain the integrality gap to verify that the solution achieved from the rounding technique is the best one. The integrality gap can be calculated by \cite{ndikumana2020deep}
\begin{equation}
\mu_i = \min_{\boldsymbol{\delta}}\frac{\tilde{\mathcal{E}}_i\left(\boldsymbol{\delta_i}^*,\boldsymbol{l_i}^*, \boldsymbol{f_i}^*,\boldsymbol{\phi_i}^*\right)}{\tilde{\mathcal{E}}_i\left(\boldsymbol{\delta_i}^*,\boldsymbol{l_i}^*, \boldsymbol{f_i}^*, \boldsymbol{\phi_i}^*\right) + \tau\Delta}, \tag{34}
\end{equation}
where $\tilde{\mathcal{E}}_i\left(\boldsymbol{\delta_i}^*,\boldsymbol{l_i}^*, \boldsymbol{f_i}^*,\boldsymbol{\phi_i}^*\right)$ is the solution obtained from the relaxed solution whereas $\tilde{\mathcal{E}}_i(\boldsymbol{\delta_i}^*,\boldsymbol{l_i}^*, \boldsymbol{f_i}^*,\boldsymbol{\phi_i}^*) + \tau\Delta$ is the solution achieved after rounding.
The best solution can be guaranteed when the value of $\mu_i$ approaches to $1$, i.e., $\mu_i \le 1$. For every relaxation, given $\tilde{\mathcal{E}}_i\left(\boldsymbol{\delta_i}^*,\boldsymbol{l_i}^*, \boldsymbol{f_i}^*,\boldsymbol{\phi_i}^*\right)$ whose instances form a convex set, the oblivious rounding scheme defined as $\tilde{\mathcal{E}}_i\left(\boldsymbol{\tilde{\delta_i}}^*,\boldsymbol{l_i}^*, \boldsymbol{f_i}^*,\boldsymbol{\phi_i}^*\right)$ is individually tight \cite{feige_et_al:LIPIcs:2016:6631}.
\begin{algorithm}[t]  
	\begin{enumerate}
		\item Initialization: Set $r=0$, $\epsilon > 0$,  and find the initial feasible points, ($\boldsymbol{\delta}^{(0)}$, $\boldsymbol{l}^{(0)}$, $\boldsymbol{f}^{(0)}$, $\boldsymbol{\phi}^{(0)}$); 
		\item \textbf{repeat}
		\item \ \ \ Choose index set $\mathcal{B}^r$;
		\item \ \ \ Let $\boldsymbol{\delta}_i^{(r+1)} \in \min_{\delta_i \in \mathcal{D}}\mathcal{E}_i(\boldsymbol{\delta}_i, \boldsymbol{\delta}^{(r)}, \boldsymbol{l}^{(r)}, \boldsymbol{f}^{(r)},\boldsymbol{\phi}^{(r)})$;
		\item \ Set $\boldsymbol{\delta}_j^{(r+1)}=\boldsymbol{\delta}_j^{(r)}, \forall j \notin \mathcal{B}^r$ and solve $\min_{\boldsymbol{\delta}_i \in \mathcal{D}}\mathcal{E}_i(\boldsymbol{\delta}_i, \boldsymbol{\delta}^{(r)}, \boldsymbol{l}^{(r)}, \boldsymbol{f}^{(r)}, \boldsymbol{\phi}^{(r)})$;
		\item \ Similarly, solve (28), (29), and (30) to obtain $\boldsymbol{l}_i^{(r+1)}, \boldsymbol{f}_i^{(r+1)}$ and $\boldsymbol{\phi}_i^{(r+1)}$ by using steps 1 through 5;
		\item \ \ \ Update $r = r + 1$;
		\item \textbf{until} $||\frac{\mathcal{E}_i^{(r)}-\mathcal{E}_i^{(r+1)}}{\mathcal{E}_i^{(r)}}|| \le \epsilon$;  
		\item Apply rounding technique on $\boldsymbol{\delta}_i^{(r+1)}$ to ensure the binary values;
        \item Solve $\tilde{\mathcal{E}}_i + \tau\Delta$ and evaluate $\mu_i$ until $\mu_i \le 1$;
		\item Finally, $\boldsymbol{\delta}^* = \boldsymbol{\delta}_i^{(r+1)}, \boldsymbol{l}^* = \boldsymbol{l}_i^{(r+1)}, \boldsymbol{f}^* = \boldsymbol{f}_i^{(r+1)},$ and $\boldsymbol{\phi}^* = \boldsymbol{\phi}_i^{(r+1)}$ are set as 
		the desired solutions.   
	\end{enumerate}
	\caption{BSUM-based Joint Resource Allocation and Task Offloading}	
	\label{algorithm:1}
\end{algorithm}

\section{Simulation Results} \label{sec: SimulationResults}
\subsection{Algorithm Design}
In this section, we present the detailed procedures of the proposed approach shown in Algorithm \ref{algorithm:1}. In the initialization step of the proposed algorithm, we determine the initial feasible points, $\left(\boldsymbol{\delta}^{(0)}, \boldsymbol{l}^{(0)}, \boldsymbol{f}^{(0)}, \boldsymbol{\phi}^{(0)}\right)$, of problem (\ref{eq26}) by setting $r=0$ and $\epsilon$ as a small positive number. Then, at each iteration $r$, the index set $\mathcal{B}^r$ is selected to begin the iterative process. The updated solution is obtained at every iteration $r+1$ by solving problems (\ref{eq27}), (\ref{eq28}), (\ref{eq29}), and (\ref{eq30}) until the convergence condition is met, i.e., $||\frac{\mathcal{E}_i^{(r)}-\mathcal{E}_i^{(r+1)}}{\mathcal{E}_i^{(r)}}|| \le \epsilon$. To enforce the solution obtained from (\ref{eq27}) to be a binary value, we apply rounding technique to it and solve $\tilde{\mathcal{E}}_i + \tau\Delta$. Finally, $\boldsymbol{\delta}_i^{(r+1)}$, $\boldsymbol{l}_i^{(r+1)}$, $\boldsymbol{f}_i^{(r+1)}$, and $\boldsymbol{\phi}_i^{(r+1)}$ are considered as the desired solutions.
\subsection{Simulation Environment}
We consider the area of interest to be $300$ m $\times$ $300$ m in which there are $5$ MEC-enabled UAVs and $30$ mobile devices. The location of the TBS is set at $(0,0,0)$. The mobile devices are randomly distributed in the considered area and the association between UAVs and MDs are determined by using k-means clustering algorithm. UAVs are assumed to be hovering at the fixed altitude of $150$ m during the considered time interval. Unless stated otherwise, the values of simulation parameters are listed in Table \ref{fig:Table_2}.

\begin{figure}[t!]
	\centering
	\includegraphics[width=\linewidth]{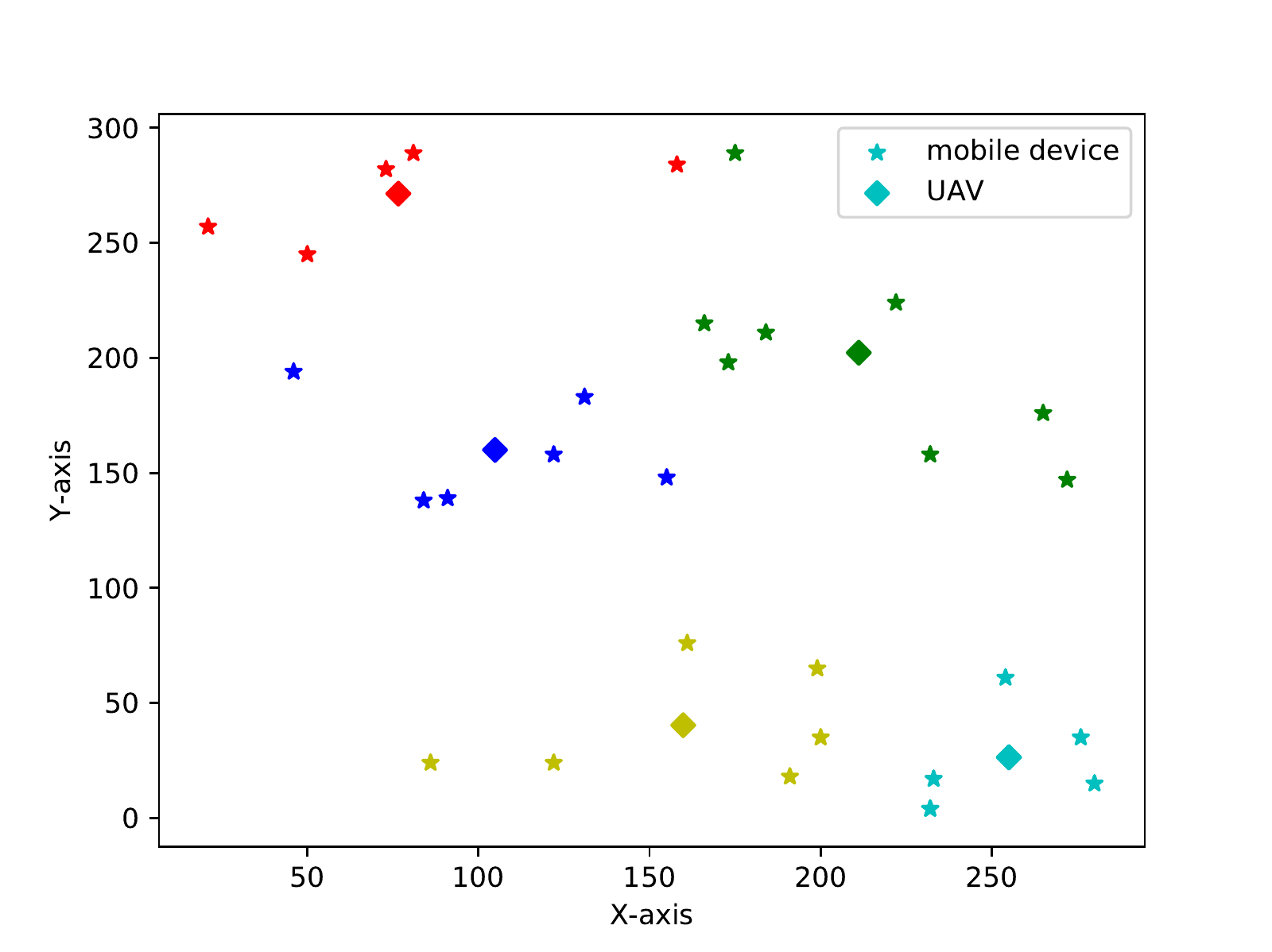}
	\caption{Association of MDs with UAVs.}
	\label{fig:device_UAV_association}
\end{figure}

In Fig. \ref{fig:device_UAV_association}, we present the association of mobile devices to UAVs by exploiting the k-means clustering algorithm. As we can see from Fig. \ref{fig:device_UAV_association} that the number of associated devices to UAV $1$, UAV $2$, UAV $3$, UAV $4$, and UAV $5$ are $8$, $6$, $5$, $6$, and $5$, respectively. Since, the association of mobile devices to UAVs is determined based on the distance, the mobile devices can experience better line-of-sight link as well as minimize the transmission energy consumption to offload their tasks.
\begin{table}[t!] 
	\caption{Simulation Parameters}
	\label{fig:Table_2}
	\begin{center}
		\begin{tabular}{|p{1.5cm}|p{2cm}||p{1.5cm}|p{2cm}|}
			\hline
			\textbf{Parameter} & \textbf{Value} &\textbf{Parameter} & \textbf{Value} \\ \hline \hline
			\ $h_0$  & $-50$ dB & $\alpha$  & $2$  \\ \hline  
			\ $P_{u,m}^n$ & $1$ mW & $P_{m,0}$ & $1$ W \\ \hline
			\ $N_0$ & $-170$ dBm &  $B$ & $20$ MHz\\ \hline
			\ $I_{u,m}$ & $[200,700]$ Mb & $O_{u,m}$ & $1000$ cycles\\ \hline
			\ $f_{u,m}$ & $[0.5,3]$ MHz & $f_m^e$ & $[1.2,2]$ GHz\\ \hline
			\ $k$  & $10^{-28}$ & $\omega$ & $180$ kHz\\ \hline
			\ $\zeta$ & $30$ N \cite{Monwar2018} & $\eta_m$ & $70\%$ \cite{Monwar2018}\\ \hline
			\ $q$ & $4$ \cite{Monwar2018} &$r$ & $0.254$ m \cite{Monwar2018} \\ \hline 
			\ $\rho$ & $1.225$ kg/m$^3$ &  $k'$  & $10^{-28}$\\ \hline   			
		\end{tabular}
	\end{center}
\end{table}
\subsection{Convergence Analysis}  

\begin{figure}[t!]
	\centering
	\includegraphics[width=\linewidth]{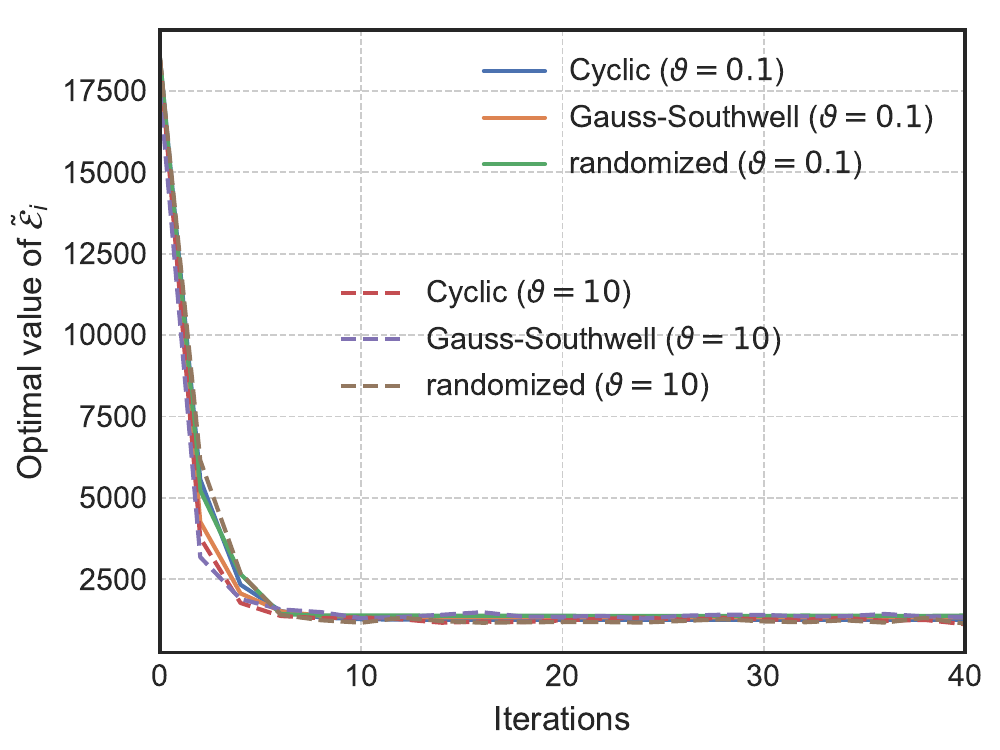}
	\caption{Convergence analysis.}
	\label{fig:convergence}
\end{figure}

In Fig. \ref{fig:convergence}, we illustrate the convergence of our proposed algorithm by applying three coordinate selection rules \cite{Hong2016}, namely, cyclic, Gauss-Southwell and randomized for two scenarios: $\vartheta = 0.1$ and $\vartheta = 10$. As we can observe from Fig. \ref{fig:convergence}, the proposed algorithm converges to a coordinate-wise minimum and stationary point at which the vectors $\boldsymbol{\delta}^* = \boldsymbol{\delta}_i^{(r+1)}, \ \boldsymbol{l}^* = \boldsymbol{l}_i^{(r+1)}, \ \boldsymbol{f}^* = \boldsymbol{f}_i^{(r+1)},$ and $\boldsymbol{\phi}^* = \boldsymbol{\phi}_i^{(r+1)}$ cannot find the better minimum direction.
\subsection{Offloaded Data Analysis}
\begin{figure}[t]
	\centering
	\includegraphics[width=\linewidth]{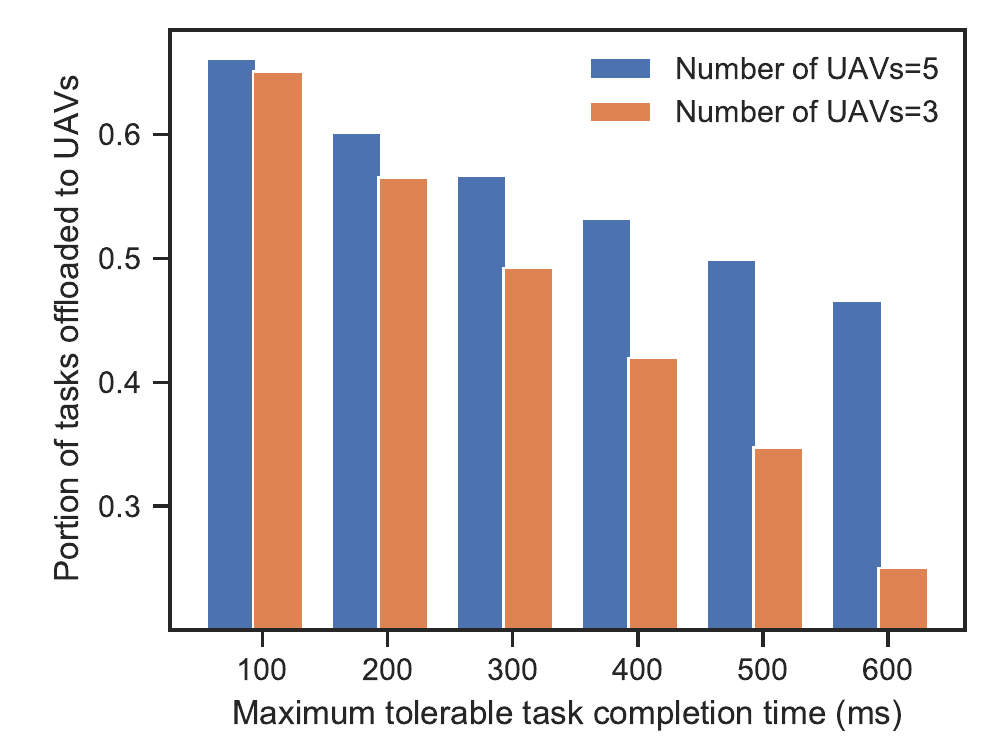}
	\caption{The impact of task completion deadline on offloaded data size of MDs to UAVs.}
	\label{fig:off_data_vs_time}
\end{figure}

The variation of the offloaded data size of mobile devices depending on the tolerable task completion deadline is illustrated in Fig. \ref{fig:off_data_vs_time}. When the devices are more tolerable to the task completion time, they will offload less task data to the UAVs so that the communication resources can be less consumed. Nevertheless, mobile devices will offload their tasks more when there are more UAVs in order to save their energy on the local computation.
\begin{figure}[h!]
	\centering
	\includegraphics[width=\linewidth]{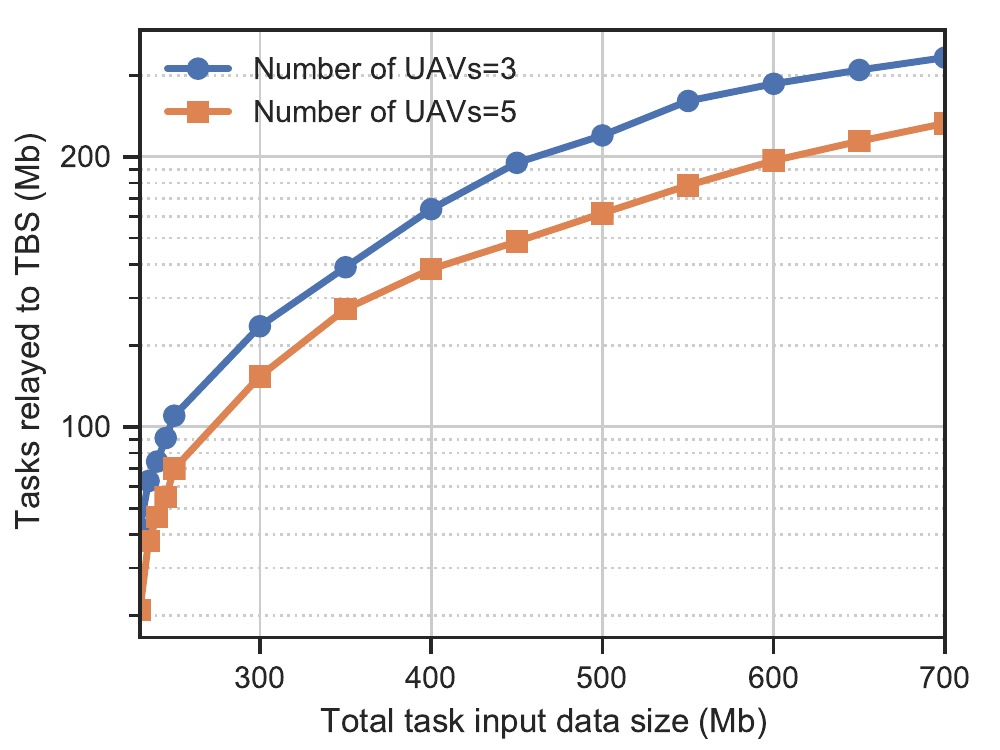}
	\caption{The impact of task input data size on the size of tasks relayed to the TBS by UAVs.}
	\label{fig:off_data_BS_vs_data_size}
\end{figure}
\begin{figure}[h!]
	\centering
	\includegraphics[width=\linewidth]{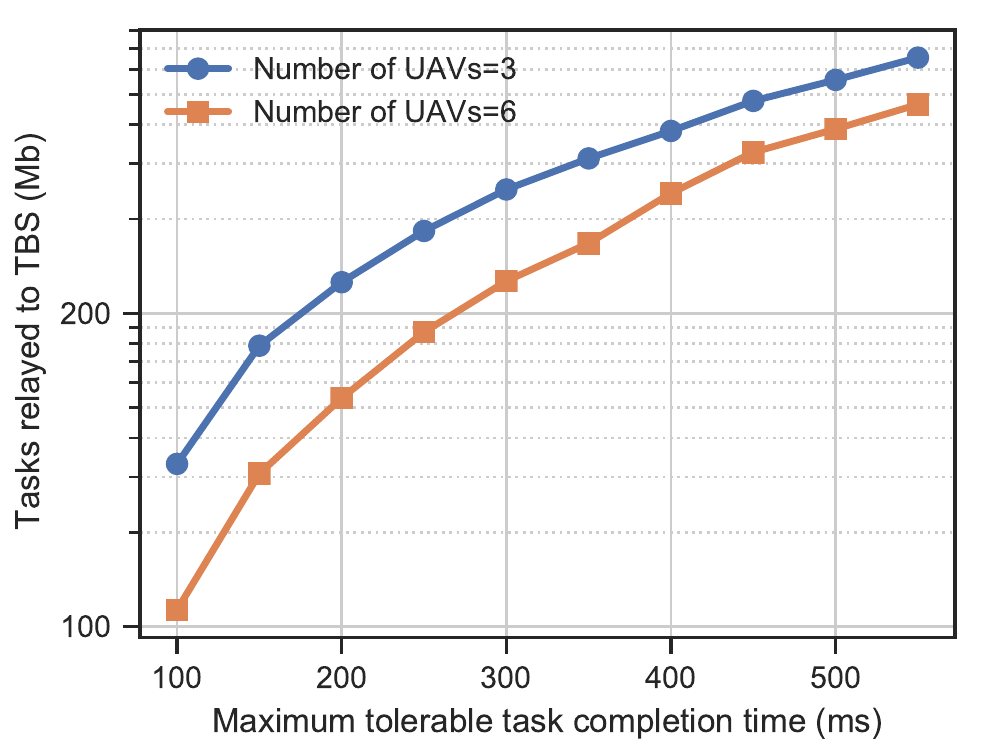}
	\caption{Tasks relayed to the TBS by UAVs vs. task completion deadline.}
	\label{fig:off_data_BS_vs_task_deadline}
\end{figure} 
\begin{figure}[h!]
	\centering
	\includegraphics[width=\linewidth]{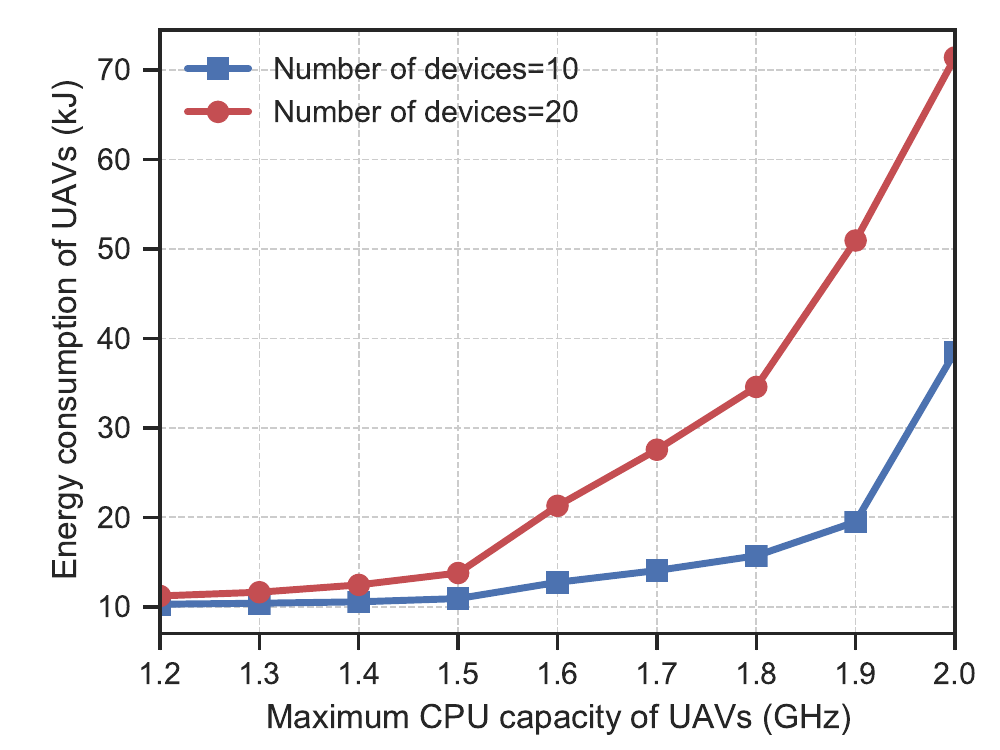}
	\caption{Energy consumption of UAVs vs. maximum CPU capacity of UAVs.}
	\label{fig:uav_energy_cpu}
\end{figure}

The data size of the task relayed to the TBS by the UAVs versus the task input data size is plotted in Fig. \ref{fig:off_data_BS_vs_data_size}. As we can see from Fig. \ref{fig:off_data_BS_vs_data_size} that the portion of task data relayed to TBS increases when the task input data size of the mobile devices increases. The reason is that the UAVs will relay more task to the TBS due to their limited CPU resources and energy budget. On the other hand, the offloaded portion to the TBS will decrease with the increasing number of UAVs. The reason is that UAVs will locally handle the offloaded tasks from the mobile devices to meet the task completion deadline constraint when they have sufficient resources to serve the associated mobile devices.

In Fig. \ref{fig:off_data_BS_vs_task_deadline}, we depict the portion of the task relayed to the TBS versus task completion deadline. When the mobile devices have a longer task completion deadline, the portion of task relayed to the TBS will increase. Due to UAVs' limited computing resources and energy budget, more task data will be relayed to the TBS by the UAVs to meet the stringent task completion deadline and save the energy consumption. When the more number of UAVs are deployed, the less task data will be relayed to the TBS. However, when the task completion deadline are more tolerable, the gap between them will become narrower.

\subsection{Energy Consumption Analysis}
In Fig. \ref{fig:uav_energy_cpu}, we depict how the amount of computing resources (CPU cycles) of UAVs impacts their energy consumption. We can observe from Fig. \ref{fig:uav_energy_cpu} that the energy consumed by the UAVs increases with the amount of CPU resources. This is because the devices tend to offload more tasks to UAVs which have rich computing resources and as a result, UAVs consume more energy on the processing of the tasks. Moreover, the number of mobile devices associated with the UAVs affects the energy consumption of the UAVs. To show that, we simulate by considering a different number of devices. It is obvious that more energy will consume when there are more number of devices to be served by the UAVs in the system.

\begin{figure}[t!]
	\centering
	\includegraphics[width=\linewidth]{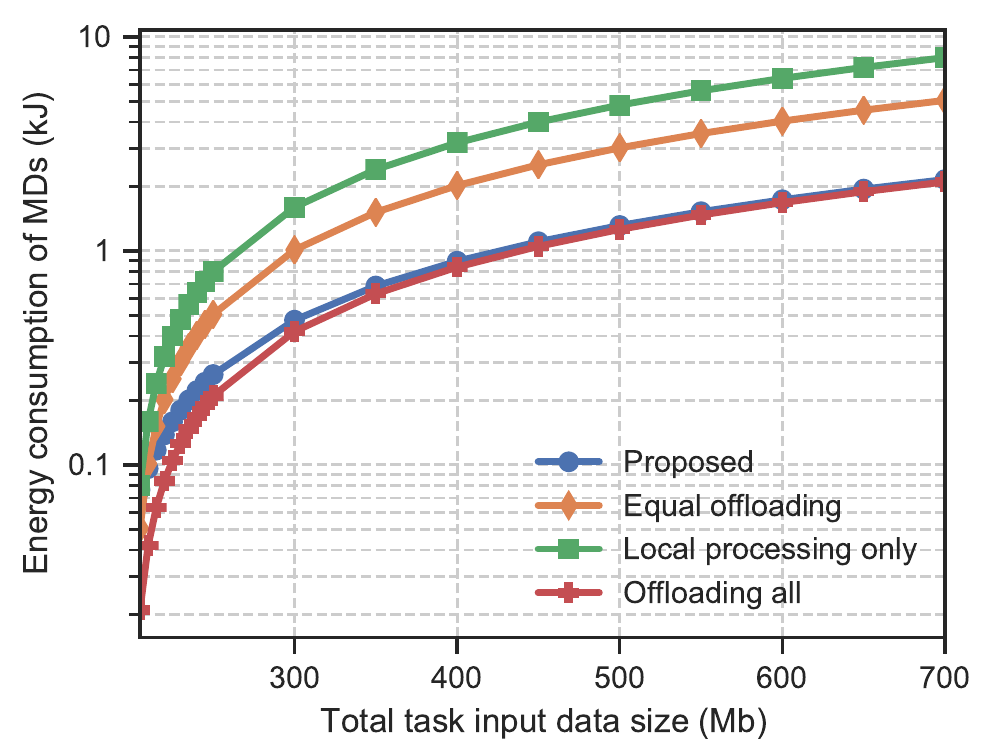}
	\caption{The impact of task input data size on energy consumption of MDs.}
	\label{fig:users_energy_vs_data}
\end{figure}

In Fig. \ref{fig:users_energy_vs_data}, we show the impact of task data size on the energy consumption of mobile devices by comparing with other benchmark schemes such as equal offloading, local processing only and offloading all. As observed in Fig. \ref{fig:users_energy_vs_data}, our proposed scheme achieves mobile devices' minimum energy consumption as in offloading all when the amount of data size increases. This is because when the mobile devices have more task input data to execute, they prefer to offload more to the UAVs to save their energy consumption. However, mobile devices consume much more energy on local computing in the equal offloading and local processing only. 

\begin{figure}[t!]
	\centering
	\includegraphics[width=\linewidth]{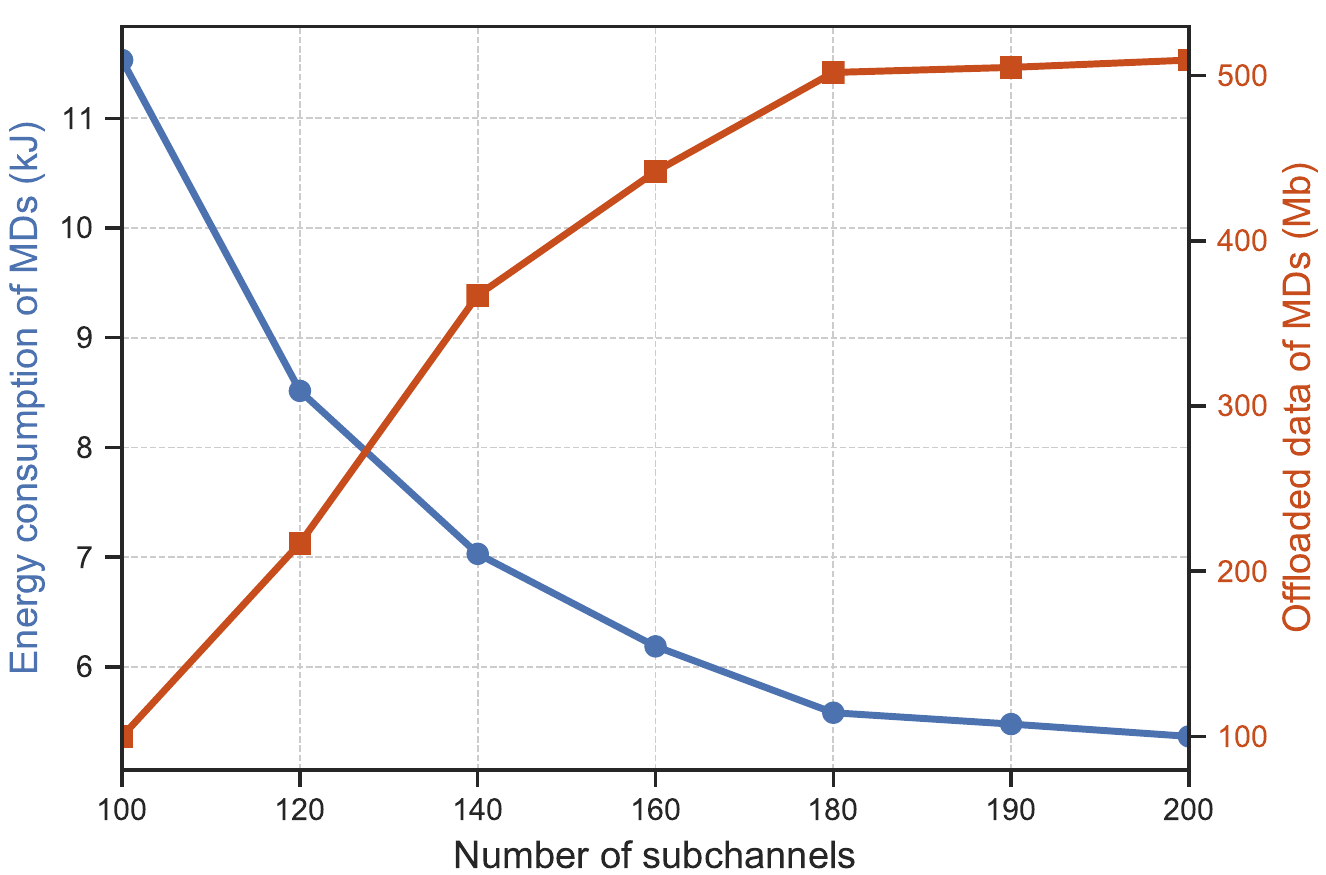}
	\caption{The impact of subchannels on MDs' energy consumption and offloaded data size.}
	\label{fig:energy_off_data_vs_subcarriers}
\end{figure}

The effects of the number of available subchannels on the devices' energy consumption and offloaded data size are given in Fig. \ref{fig:energy_off_data_vs_subcarriers}. As we can observe from Fig. \ref{fig:energy_off_data_vs_subcarriers}, the energy dissipated by the mobile devices reduces with the increasing number of subchannels. This is because mobile devices can minimize their transmit power by selecting the more favorable subchannel while offloading their tasks to the UAVs. On the other hand, they can offload more data to the UAVs by saving energy consumption on data transmission.

\begin{figure}[t!]
	\centering
	\includegraphics[width=\linewidth]{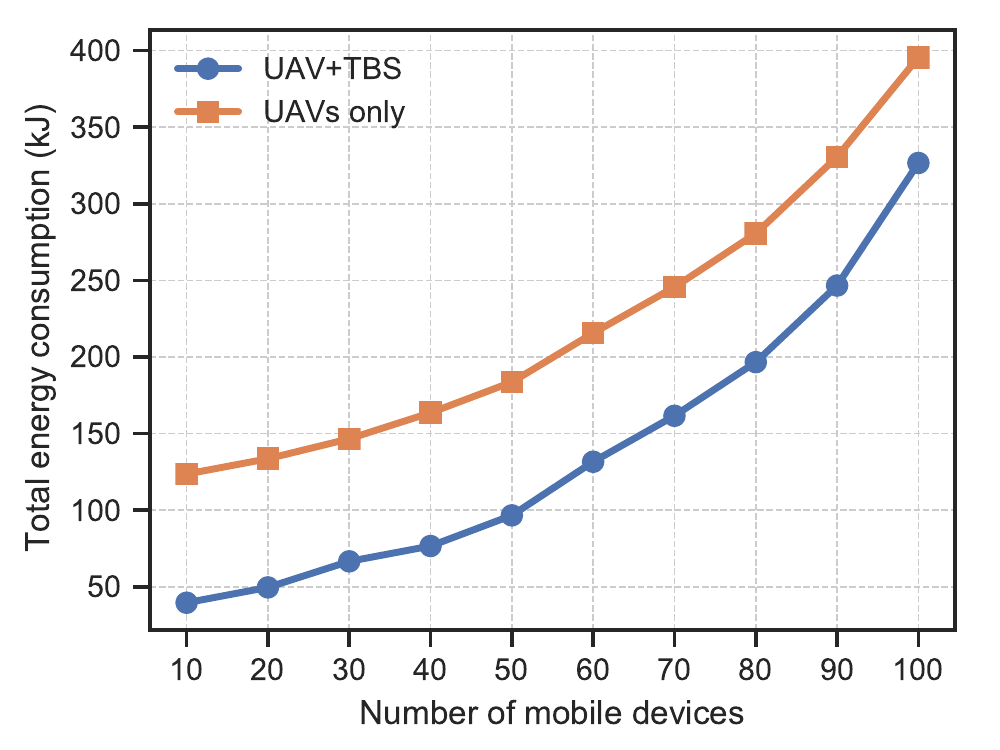}
	\caption{Total energy consumption of the system vs number of mobile devices.}
	\label{fig:energy_vs_users}
\end{figure}

In Fig. \ref{fig:energy_vs_users}, we study the effectiveness of our proposed approach in terms of the system's total energy consumption by varying the number of mobile devices. The total energy consumption of the system increases with the increasing number of devices. We compare our proposed scheme (UAV+TBS) with the other scheme (UAVs only) to verify how the proposed scheme can give better results in terms of total energy consumption. The total energy consumption while considering UAVs only is higher than that UAVs and TBS collaboration. The reason is that both the mobile devices and UAVs might consume too much energy on the task execution without the assistance of the TBS.

\section{Conclusion} \label{sec: Conclusion}
In this paper, we have studied a multi-UAV-assisted two-stage MEC system in which MEC-enabled UAVs provide computing and relaying services to the mobile devices. Taking into account the tolerable delay of the tasks and the limited communication/computation resources of the UAVs, we have formulated a joint resource allocation and offloading problem with the objective of minimizing the total energy consumption of the mobile devices and UAVs. Since the formulated optimization problem is a mixed-integer non-convex problem which is NP-hard, we first relax the channel assignment variable and reformulated the problem. However, the reformulated problem is still non-convex due to the coupling among the variables. To address that problem, the BSUM algorithm has been deployed. The simulation results have shown that the proposed approach can reduce the energy consumption of the network and outperformed the baseline schemes.

\bibliographystyle{IEEEtran}
\bibliography{Multi_UAV_reference}

\begin{thebibliography}{10}
\providecommand{\url}[1]{#1}
\csname url@samestyle\endcsname
\providecommand{\newblock}{\relax}
\providecommand{\bibinfo}[2]{#2}
\providecommand{\BIBentrySTDinterwordspacing}{\spaceskip=0pt\relax}
\providecommand{\BIBentryALTinterwordstretchfactor}{4}
\providecommand{\BIBentryALTinterwordspacing}{\spaceskip=\fontdimen2\font plus
\BIBentryALTinterwordstretchfactor\fontdimen3\font minus
  \fontdimen4\font\relax}
\providecommand{\BIBforeignlanguage}[2]{{%
\expandafter\ifx\csname l@#1\endcsname\relax
\typeout{** WARNING: IEEEtran.bst: No hyphenation pattern has been}%
\typeout{** loaded for the language `#1'. Using the pattern for}%
\typeout{** the default language instead.}%
\else
\language=\csname l@#1\endcsname
\fi
#2}}
\providecommand{\BIBdecl}{\relax}
\BIBdecl

\bibitem{Mao2017}
Y.~Mao, C.~You, J.~Zhang, K.~Huang, and K.~B. Letaief, ``A survey on mobile
  edge computing: The communication perspective,'' \emph{{IEEE} Communications
  Surveys {\&} Tutorials}, vol.~19, no.~4, pp. 2322--2358, Aug. 2017.

\bibitem{Hu2019}
X.~Hu, K.-K. Wong, K.~Yang, and Z.~Zheng, ``{UAV}-assisted relaying and edge
  computing: Scheduling and trajectory optimization,'' \emph{{IEEE}
  Transactions on Wireless Communications}, vol.~18, no.~10, pp. 4738--4752,
  Oct. 2019.

\bibitem{zeng2019accessing}
Y.~Zeng, Q.~Wu, and R.~Zhang, ``Accessing from the sky: A tutorial on uav
  communications for 5{G} and beyond,'' \emph{Proc. of the IEEE}, vol. 107,
  no.~12, pp. 2327--2375, Dec. 2019.

\bibitem{zeng2016wireless}
Y.~Zeng, R.~Zhang, and T.~J. Lim, ``Wireless communications with unmanned
  aerial vehicles: opportunities and challenges,'' \emph{IEEE Communications
  Magazine}, vol.~54, no.~5, pp. 36--42, May 2016.

\bibitem{Mozaffari2016a}
M.~Mozaffari, W.~Saad, M.~Bennis, and M.~Debbah, ``Unmanned aerial vehicle with
  underlaid device-to-device communications: Performance and tradeoffs,''
  \emph{{IEEE} Transactions on Wireless Communications}, vol.~15, no.~6, pp.
  3949--3963, Jun. 2016.

\bibitem{mozaffari2016efficient}
------, ``Efficient deployment of multiple unmanned aerial vehicles for optimal
  wireless coverage,'' \emph{IEEE Communications Letters}, vol.~20, no.~8, pp.
  1647--1650, Jun. 2016.

\bibitem{Sharma2016}
V.~Sharma, M.~Bennis, and R.~Kumar, ``{UAV}-assisted heterogeneous networks for
  capacity enhancement,'' \emph{{IEEE} Communications Letters}, vol.~20, no.~6,
  pp. 1207--1210, Jun. 2016.

\bibitem{mozaffari2017wireless}
M.~Mozaffari, W.~Saad, M.~Bennis, and M.Debbah, ``Wireless communication using
  unmanned aerial vehicles ({UAVs}): Optimal transport theory for hover time
  optimization,'' \emph{IEEE Transactions on Wireless Communications}, vol.~16,
  no.~12, pp. 8052--8066, Sep. 2017.

\bibitem{Mozaffari2016}
M.~Mozaffari, W.~Saad, M.~Bennis, and M.~Debbah, ``Optimal transport theory for
  power-efficient deployment of unmanned aerial vehicles,'' in \emph{2016
  {IEEE} International Conference on Communications ({ICC})}.\hskip 1em plus
  0.5em minus 0.4em\relax {IEEE}, Kuala Lumpur, Malaysia, May. 2016.

\bibitem{Mozaffari2019}
M.~Mozaffari, A.~T.~Z. Kasgari, W.~Saad, M.~Bennis, and M.~Debbah, ``Beyond
  {5G} with {UAVs}: Foundations of a 3{D} wireless cellular network,''
  \emph{{IEEE} Transactions on Wireless Communications}, vol.~18, no.~1, pp.
  357--372, Jan. 2019.

\bibitem{Zhao2018}
H.~Zhao, H.~Wang, W.~Wu, and J.~Wei, ``Deployment algorithms for {UAV} airborne
  networks toward on-demand coverage,'' \emph{{IEEE} Journal on Selected Areas
  in Communications}, vol.~36, no.~9, pp. 2015--2031, Sep. 2018.

\bibitem{Mozaffari2017}
M.~Mozaffari, W.~Saad, M.~Bennis, and M.~Debbah, ``Mobile unmanned aerial
  vehicles ({UAVs}) for energy-efficient internet of things communications,''
  \emph{{IEEE} Transactions on Wireless Communications}, vol.~16, no.~11, pp.
  7574--7589, Nov. 2017.

\bibitem{Ei2019}
N.~N. Ei, C.~W. Zaw, M.~K. Lee, and C.~S. Hong, ``Cell association in
  energy-constrained unmanned aerial vehicle communications under altitude
  consideration,'' in \emph{{IEEE} International Conference on Information
  Networking ({ICOIN})}, Kuala Lumpur, Malaysia, Jan. 2019.

\bibitem{Zhang2019}
S.~Zhang, H.~Zhang, B.~Di, and L.~Song, ``Cellular {UAV}-to-{X} communications:
  Design and optimization for multi-{UAV} networks,'' \emph{{IEEE} Transactions
  on Wireless Communications}, vol.~18, no.~2, pp. 1346--1359, Feb. 2019.

\bibitem{Chen2017}
M.~Chen, M.~Mozaffari, W.~Saad, C.~Yin, M.~Debbah, and C.~S. Hong, ``Caching in
  the sky: Proactive deployment of cache-enabled unmanned aerial vehicles for
  optimized quality-of-experience,'' \emph{{IEEE} Journal on Selected Areas in
  Communications}, vol.~35, no.~5, pp. 1046--1061, May 2017.

\bibitem{Ning2019}
Z.~Ning, P.~Dong, X.~Kong, and F.~Xia, ``A cooperative partial computation
  offloading scheme for mobile edge computing enabled internet of things,''
  \emph{{IEEE} Internet of Things Journal}, vol.~6, no.~3, pp. 4804--4814, Jun.
  2019.

\bibitem{Ren2017}
J.~Ren, G.~Yu, Y.~Cai, Y.~He, and F.~Qu, ``Partial offloading for latency
  minimization in mobile-edge computing,'' in \emph{{IEEE} Global
  Communications Conference}, Singapore, Dec. 2017.

\bibitem{Liu2017}
C.-F. Liu, M.~Bennis, and H.~V. Poor, ``Latency and reliability-aware task
  offloading and resource allocation for mobile edge computing,'' in
  \emph{{IEEE} Globecom Workshops ({GC} Wkshps)}, Singapore, Dec. 2017.

\bibitem{gao2020computation}
M.~Gao, R.~Shen, J.~Li, S.~Yan, Y.~Li, J.~Shi, Z.~Han, and L.~Zhuo,
  ``Computation offloading with instantaneous load billing for mobile edge
  computing,'' \emph{IEEE Transactions on Services Computing (Early Access)},
  May 2020.

\bibitem{Chen2020}
X.~Chen, T.~Chen, Z.~Zhao, H.~Zhang, M.~Bennis, and Y.~JI, ``Resource awareness
  in unmanned aerial vehicle-assisted mobile-edge computing systems,'' in
  \emph{{IEEE} 91st Vehicular Technology Conference ({VTC}2020-Spring)},
  Antwerp, Belgium, May 2020.

\bibitem{tun2020energy}
Y.~K. Tun, Y.~M. Park, N.~H. Tran, W.~Saad, S.~R. Pandey, and C.~S. Hong,
  ``Energy-efficient resource management in {UAV}-assisted mobile edge
  computing,'' \emph{IEEE Communications Letters (Early Access)}, Sep. 2020.

\bibitem{Alsenwi2020}
M.~Alsenwi, Y.~K. Tun, S.~R. Pandey, N.~N. Ei, and C.~S. Hong, ``{UAV}-assisted
  multi-access edge computing system: An energy-efficient resource management
  framework,'' in \emph{{IEEE} International Conference on Information
  Networking ({ICOIN})}, Barcelona, Spain, Jan. 2020.

\bibitem{Zhang2019a}
J.~Zhang, L.~Zhou, Q.~Tang, E.~C.-H. Ngai, X.~Hu, H.~Zhao, and J.~Wei,
  ``Stochastic computation offloading and trajectory scheduling for
  {UAV}-assisted mobile edge computing,'' \emph{{IEEE} Internet of Things
  Journal}, vol.~6, no.~2, pp. 3688--3699, Apr. 2019.

\bibitem{Li2020}
M.~Li, N.~Cheng, J.~Gao, Y.~Wang, L.~Zhao, and X.~Shen, ``Energy-efficient
  {UAV}-assisted mobile edge computing: Resource allocation and trajectory
  optimization,'' \emph{{IEEE} Transactions on Vehicular Technology}, vol.~69,
  no.~3, pp. 3424--3438, Mar. 2020.

\bibitem{Zhang2020}
Q.~Zhang, J.~Chen, L.~Ji, Z.~Feng, Z.~Han, and Z.~Chen, ``Response delay
  optimization in mobile edge computing enabled {UAV} swarm,'' \emph{{IEEE}
  Transactions on Vehicular Technology}, vol.~69, no.~3, pp. 3280--3295, Mar.
  2020.

\bibitem{wu2018common}
Q.~Wu and R.~Zhang, ``Common throughput maximization in {UAV}-enabled {OFDMA}
  systems with delay consideration,'' \emph{IEEE Transactions on
  Communications}, vol.~66, no.~12, pp. 6614--6627, Aug. 2018.

\bibitem{Al-Hourani2014}
A.~Al-Hourani, S.~Kandeepan, and A.~Jamalipour, ``Modeling air-to-ground path
  loss for low altitude platforms in urban environments,'' in \emph{{IEEE}
  Global Communications Conference}, Austin, TX, USA, Dec. 2014.

\bibitem{Monwar2018}
M.~Monwar, O.~Semiari, and W.~Saad, ``Optimized path planning for inspection by
  unmanned aerial vehicles swarm with energy constraints,'' in \emph{{IEEE}
  Global Communications Conference ({GLOBECOM})}, Abu Dhabi, United Arab
  Emirates, Dec. 2018.

\bibitem{Stolaroff2018}
J.~K. Stolaroff, C.~Samaras, E.~R. O'Neill, A.~Lubers, A.~S. Mitchell, and
  D.~Ceperley, ``Energy use and life cycle greenhouse gas emissions of drones
  for commercial package delivery,'' \emph{Nature Communications}, vol.~9,
  no.~1, pp. 1--13, Feb. 2018.

\bibitem{Hong2016a}
Y.-W.~P. Hong, T.-C. Hsu, and P.~Chennakesavula, ``Wireless power transfer for
  distributed estimation in wireless passive sensor networks,'' \emph{{IEEE}
  Transactions on Signal Processing}, vol.~64, no.~20, pp. 5382--5395, Oct.
  2016.

\bibitem{Hong2016}
M.~Hong, M.~Razaviyayn, Z.-Q. Luo, and J.-S. Pang, ``A unified algorithmic
  framework for block-structured optimization involving big data: With
  applications in machine learning and signal processing,'' \emph{{IEEE} Signal
  Processing Magazine}, vol.~33, no.~1, pp. 57--77, Jan. 2016.

\bibitem{Ndikumana2020a}
A.~Ndikumana, N.~H. Tran, T.~M. Ho, Z.~Han, W.~Saad, D.~Niyato, and C.~S. Hong,
  ``Joint communication, computation, caching, and control in big data
  multi-access edge computing,'' \emph{{IEEE} Transactions on Mobile
  Computing}, vol.~19, no.~6, pp. 1359--1374, Jun. 2020.

\bibitem{Boyd2004}
S.~Boyd and L.~Vandenberghe, \emph{Convex Optimization, Cambridge University
  Press}, Mar. 2004.

\bibitem{feige_et_al:LIPIcs:2016:6631}
\BIBentryALTinterwordspacing
U.~Feige, M.~Feldman, and I.~Talgam-Cohen, ``{Oblivious Rounding and the
  Integrality Gap},'' in \emph{Approximation, Randomization, and Combinatorial
  Optimization. Algorithms and Techniques (APPROX/RANDOM 2016)}, ser. Leibniz
  International Proceedings in Informatics (LIPIcs), K.~Jansen, C.~Mathieu,
  J.~D.~P. Rolim, and C.~Umans, Eds., vol.~60.\hskip 1em plus 0.5em minus
  0.4em\relax Dagstuhl, Germany: Schloss Dagstuhl--Leibniz-Zentrum fuer
  Informatik, 2016, pp. 8:1--8:23. [Online]. Available:
  \url{http://drops.dagstuhl.de/opus/volltexte/2016/6631}
\BIBentrySTDinterwordspacing

\bibitem{ndikumana2020deep}
A.~Ndikumana, N.~H. Tran, K.~T. Kim, C.~S. Hong \emph{et~al.}, ``Deep learning
  based caching for self-driving cars in multi-access edge computing,''
  \emph{IEEE Transactions on Intelligent Transportation Systems (Early
  Access)}, Mar. 2020.

\end{thebibliography}
\end{document}